# Automated and optimally FRET-assisted structural modeling

Mykola Dimura[1, 2], Thomas O. Peulen[1], Hugo Sanabria[1, 3], Dmitro Rodnin[1], Katherina Hemmen[1], Claus A.M. Seidel[1, *], Holger Gohlke[2, 4, *]

[1] Chair for Molecular Physical Chemistry, Heinrich Heine University Düsseldorf, 40225 Düsseldorf, Germany;
[2] Institute for Pharmaceutical and Medicinal Chemistry, Heinrich Heine University Düsseldorf, 40225 Düsseldorf, Germany;
[3] Department of Physics and Astronomy, Clemson University, Clemson, South Carolina, U.S.A.
[4] John von Neumann Institute for Computing (NIC), Jülich Supercomputing Centre (JSC), and Institute for Complex Systems - Structural Biochemistry (ICS-6), Forschungszentrum Jülich GmbH, 52425 Jülich, Germany

**Abstract**

FRET experiments can yield state-specific structural information on complex dynamic biomolecular assemblies. However, FRET experiments need to be combined with computer simulations to overcome their sparsity. We introduce (i) an automated FRET experiment design tool determining optimal FRET pairs for structural modeling, (ii) a protocol for efficient FRET-assisted computational structural modeling at multiple scales, and (iii) a quantitative quality estimate for judging the *accuracy* of determined structures. We tested against simulated and experimental data.

**Main text**

Structures of biomacromolecules and their complexes are often key to understanding the molecules' functions and underlying mechanisms, and therefore can be a prerequisite for related biological and medical developments. For certain classes of systems, including multi-domain proteins, biomacromolecular complexes, dynamic systems with unstructured regions, and systems with lowly populated conformational states, experimental structure determination is challenging. For such complex systems, contemporary computational structure prediction tools[1-5] often yield several alternative models, which may contain different domain folds and supertertiary structures, particularly if template structures of homologous proteins are not available. FRET experiments can alleviate these difficulties in that they yield state-specific structural information on complex constructs, even for very dynamic systems with short-lived states in the microsecond time scale[6-10]. However, FRET experiments need to be combined with computer simulations to solve the issue that FRET data is usually too sparse to cover all structural details[11,12]. The problem of quantitative *accuracy* evaluation (as opposed to *precision*) remained largely unaddressed as well. Here, we introduce (i) methodological developments for an automated design of FRET experiments that aim at obtaining the most informative set of FRET pairs optimal for structural modeling, and (ii) a protocol for efficient structure determination and quantitative quality estimation based on such FRET data and computational modeling at multiple scales.

We devised an iterative workflow for FRET-assisted modeling consisting of six steps (**Fig. 1a**), and developed the related software (see Code Availability below): (1) collection of prior knowledge, (2) generation of an initial structural ensemble, (3) selection of the most informative FRET pairs,

(4) acquisition and analysis of the experimental data, (5) FRET screening (a statistical quality assessment using a $\chi_n^2$ criterion (**Online Methods eq. 8**), (6) FRET-guided structural sampling. This workflow is exemplified for the *E. coli* YaaA protein (YaaA, **Fig. 1b-h**).

In step 1, prior information is obtained from structures in the PDB of other states of a given target, homology models, or structural models built with other computational structure prediction tools[1-5] (**Fig. 1b**). In step 2, this initial structural ensemble is expanded by conformational sampling (**Fig. 1c**). For this, multiple unrestrained simulations using structures obtained in the first step as seeds are performed using the NMSim approach, which performs normal mode-based geometric simulations for multiscale modeling of protein conformational changes[13] (http://nmsim.de). For YaaA, *prior* structures were taken from the computational structure predictions submitted to the CASP 11 experiment (T806)[14] (**Online Methods section 1**). For the other proteins, seed structures corresponding to conformational states different from the 'true' one were taken from the PDB (**Table 4**). In step 3, we use a novel algorithm for experiment planning to automatically determine a set of most informative FRET pairs (**Fig. 1d, Supplementary Fig. 1, Supplementary Table 1**) optimized for highest model precision that is based on a given *prior* structural ensemble. Additionally, our tool for experimental design can consider user-specified labeling site accessibility, chemical nature, and influence on function and stability as determined from mutation analysis or sequence coevolution data (**Online Methods section 4**). A higher number of measured FRET pairs results in higher precision, since less diverse structures are found for the ensemble within the confidence level of 68% corresponding to $\chi_n^2 < 1$ (**Fig. 1e**). This agrees with the predictions from the pair selection algorithm (**Fig. 1f**). Notably, for sparser and smaller prior ensembles less FRET measurements are needed to achieve a target precision. In step 4, FRET data is acquired including uncertainty estimates (**Supplementary Table 1**).

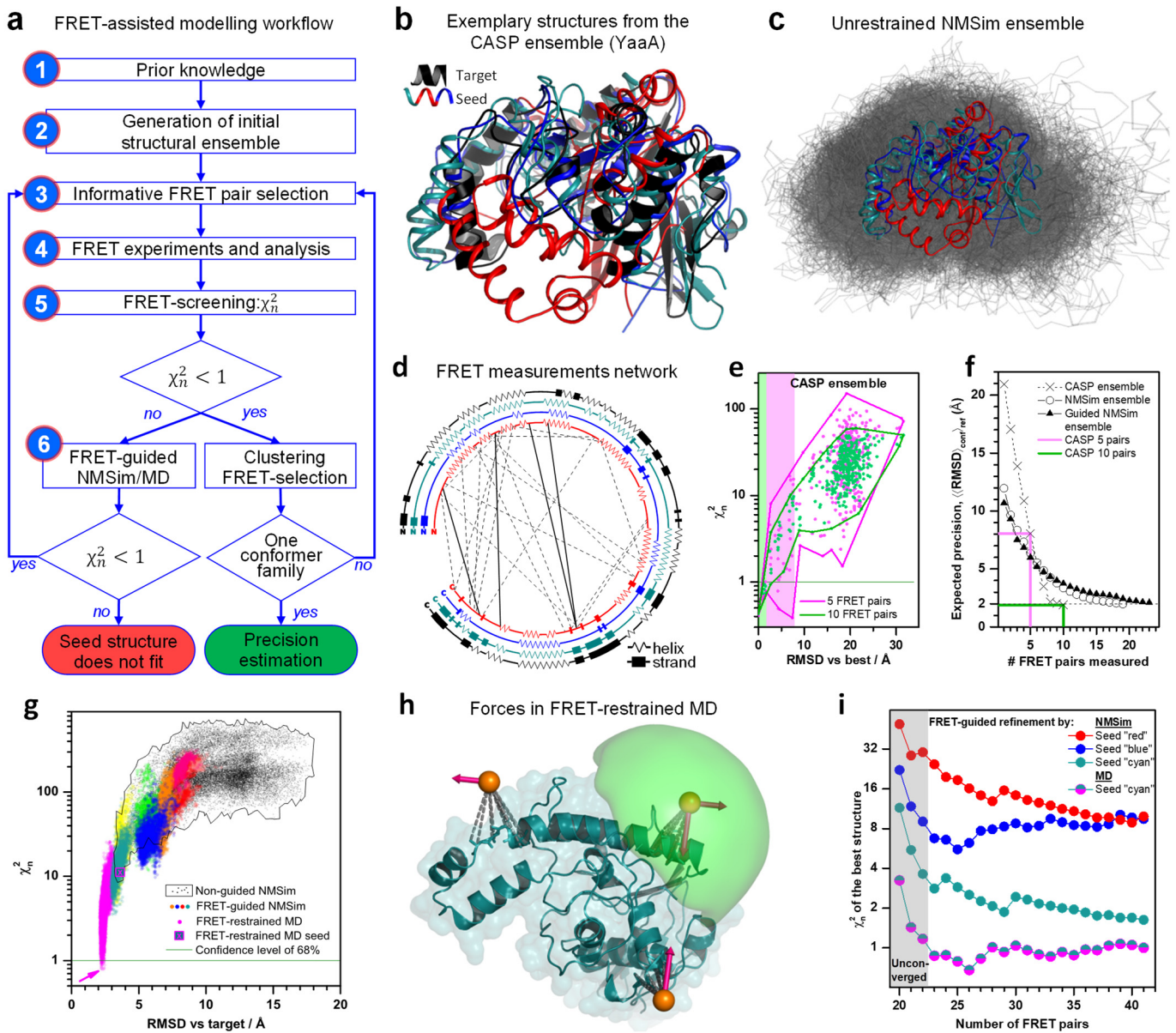

***Figure 1 | Automated FRET-assisted structure prediction on the example of the E. coli protein YaaA.***
(***a***) *Step-by-step workflow for automated and optimally FRET-assisted structural modeling.* (***b***) *Collection of prior information: CASP predictions used as seed structures (red, cyan, blue), and the target crystal structure of YaaA (black, PDB ID: 5CAJ). Three out of eleven used seed structures are shown for clarity.* (***c***) *Generation of initial structural ensemble (gray) by NMSim without any FRET information, using CASP predictions (red, cyan, blue) as seed structures.* (***d***) *Network of FRET pairs used for guided NMSim simulations (dashed) and FRET screening (dashed and solid). Secondary structure elements (zigzag – α-helix, $3_{10}$-helix or π-helix; rectangle – β-bridge or β-ladder, line – loop) for three shown seed structures (red, cyan, blue) and the target (black).* (***e***) *Impact of the number of selected FRET pairs on the precision of the selected ensemble. The $\chi_n^2$ values and RMSD against the best structure for the structural ensemble of CASP targets are shown. The diversity of the structures with lower $\chi_n^2$ defines the precision of the FRET-selected structure. The green and magenta shaded areas correspond to 10 and 5 FRET measurements,* *respectively. (**f**) Expected precision of the resulting structural model, depending on the number of used FRET measurements. For the sparser conformational ensemble (CASP ensemble, crosses), the decay is steeper than for the more diverse ensembles generated by NMSim (circles). (**g**) FRET $\chi_n^2$ values and RMSD against the crystal structure (target) for different conformations (points). Black points stand for unrestrained NMSim sampling starting from homology models. Colored points represent FRET-guided NMSim simulations. Magenta points represent FRET-restrained MD simulations. Guided simulations stemming from different homology models are shown in different colors. (**h**) Attachment (dashed gray) of pseudo-atoms (orange spheres) and application of FRET-restraints (pink arrows) in FRET-restrained MD simulations. The accessible volume of a fluorophore is shown as green surface. (**i**) $\chi_n^2$ of the best conformers generated by FRET-guided NMSim or FRET-restrained MD simulations using different seed structures. $\chi_n^2$ starts to converge with ~23 selected FRET pairs. Line colors correspond to structure colors in (**b**).*

In step 5 we screen our large ensemble to find those structural models which agree well with the FRET observables corresponding to the 'true' structure. To our knowledge, no absolute quality measure exists for this purpose so far. Thus, we introduced a quantitative and reliable accuracy estimation by computing the goodness-of-fit, $\chi_n^2 = \chi^2/\chi_{68\%}^2$, to judge the agreement (**see Online Methods, eq. 8**). $\chi_n^2$ is an absolute measurement of quality, it relies on an accurate error estimation and requires FRET measurements that have not been used for previous optimization steps. Therefore, $\chi_n^2$ is analogous to cross-validation of the structural model and similar in spirit to $R_{free}$ known from X-ray crystallography[15]. For calculating $\chi_n^2$, we introduce for the first time a tool for the automatic quantitative estimation of the number of relevant degrees of freedom in FRET-based models. The tool can be applied to an arbitrary ensemble of structural models, which opens a convenient interface for integration with third party structural modeling tools. Using the $\chi_n^2$ criterion ($\chi_n^2 <$ 1), the FRET data allow us to extract a set of conformers (**Supplementary Fig. 2**): conformations with $\chi_n^2$ values < 1.0 are identified as FRET-consistent models. If the diversity within the FRET-selected ensemble is sufficiently low (e.g., root mean square deviation ($RMSD_{ij}$) < 3 Å), the workflow is considered to converge. The diversity within the FRET-selected ensemble represents the precision of the obtained model.

However, if no structure with good FRET agreement ($\chi_n^2 < 1$) were found in the initial ensemble (**Fig. 1g**, black points), we establish two new multi-scale structural sampling tools to extend this ensemble by FRET-guided structural sampling (step (6): FRET-guided normal mode-based geometric simulations (NMSim approach[13], **Supplementary Fig. 3**) employing a Metropolis-Hastings Monte Carlo algorithm and FRET-restrained molecular dynamics simulations (**Fig. 1h**, **Supplementary Fig. 4**), which implement a novel implicit dye representation and experiment-based inter-dye distance restraints, rather than inaccurate atom-atom distance restraints. The additional FRET information allows us to explore areas of phase space inaccessible for purely computational multi-scale simulations, so that novel and experimentally relevant (super-) tertiary structures can be resolved. Strikingly, FRET-guided refinement of different seed structures yields distinct limiting $\chi_n^2$ levels for the final structural models (**Fig. 1g,i**) with more accurate folds indicated by lower $\chi_n^2$ values. This allows us to detect errors in the folds of seed conformers that cannot be easily corrected, down to the level of secondary structure (**Fig. 1d**). Note that only four additional FRET-pairs are needed here for reaching a converged $\chi_n^2$ (grey box, **Fig. 1i**).

The workflow was benchmarked on simulated and experimental data. For that, we used an exemplary set of six proteins that are diverse in their structures, sizes (148 to 409 amino acids), and types of internal interconversion motions (hinge-bending, shear, and twist), and mode of interaction (induced fit or conformational selection [16,17], **Supplementary Note 1**). Some of these proteins have been used previously to investigate conformational sampling techniques[18-20]. For each protein, at least one conformation is available in the PDB. This conformation is used as a 'true' reference structure for accuracy estimation. For five proteins realistic FRET data was simulated as described previously[8] (**Supplementary Table 1, Supplementary Fig. 5, Supplementary Note 5**). For T4 lysozyme (T4L) a comprehensive experimental data set was acquired in solution, which allowed us to resolve two short (4 μs) lived conformers referred to as "C1" and "C2"[21], which were also observed by X-ray crystallography. Using simulated and experimental datasets, we applied our FRET-guided structural modeling procedure in order to arrive at a target structural model, starting from the seed conformer corresponding to the other state. In this benchmark study, we obtained state-specific structural models with a precision of 2 to 3.5 Å and an accuracy against the target structure between 2 and 3 Å (**Table 4**, **Fig. 2, Supplementary Fig. 11**) for as few as 13 to 23 FRET measurements, depending on the structural diversity and accuracy of the prior ensemble. This parsimony is attributed to the novel method for automatic determination of a set of optimal FRET pairs (**Supplementary Fig. 1**). These results illustrate that the predictive power and reliability of $\chi_n^2$ (**Supplementary Fig. 6**) yields target structures with an observed structural heterogeneity for protein backbone conformations at room temperature as found in all-atom MD simulations and NMR experiments[22]. The resolution of experimental FRET studies is sufficient to distinguish between the known conformers C1 and C2 (**Supplementary Fig. 5**) which differ by 4 Å RMSD.

*Table 4* | *Summary over the proteins used in the benchmark*[(\*\*)].

| Protein name | PDB ID | | | RMSD / Å | | | | #pairs | |
|---|---|---|---|---|---|---|---|---|---|
| | seed | target | #aa | prior | best | min | max | guiding | +validation |
| ***E. coli* YaaA protein** | [(\*)] | 5caj | 256 | 4.7-14.6 | 2.4 | 2.2 | 2.5 | 19 | +4 |
| **LAO binding protein** | 2lao | 1lst | 238 | 4.7 | 2.4 | 1.8 | 2.4 | 12 | +3 |
| **Calmodulin** | 1cfd | 1ckk | 148 | 9.8 | 2.4 | 2.4 | 3.1 | 13 | +9 |
| **Atlastin1** | 4idn | 3q5e | 409 | 18.7 | 2.5 | 2.4 | 3.0 | 10 | +9 |
| **Adenylate kinase** | 4ake | 1ake | 214 | 7.2 | 2.3 | 2.1 | 3.2 | 10 | +8 |
| <u>**T4 lysozyme (C2→C1)**</u> | 3gun | 172l | 162 | 4.0 | 2.8 | 2.8 | 3.3 | 10 | +10 |
| <u>**T4 lysozyme (C1→C2)**</u> | 172l | 3gun | 162 | 4.0 | 2.5 | 2.0 | 3.5 | 10 | +10 |

#aa stands for the no. of amino acids in the protein, as used in the benchmark. The RMSD of the seed structure against the target structure is indicated as $RMSD_{prior}$. $RMSD_{min/best/max}$ of the FRET-selected structures against the target structure are indicated as an accuracy measure for the obtained ensembles; $RMSD_{best}$ represents the deviation for the model with the lowest $\chi_n^2$; $RMSD_{min}$ and $RMSD_{max}$ correspond to the minimum and maximum RMSD of the structural models within the confidence level. All RMSDs are calculated for $C_\alpha$ atoms only. For the T4L (underlined) experimental FRET data was used, for other proteins the data was simulated.
[(\*)]For *E. coli* YaaA protein, 10 seed structures were selected among the predictions submitted for the CASP 11 experiment (target T806). This selection differs from the target crystal structure (RMSD of 4.7 to 14.6 Å) and represents different folds and secondary structures. The number of FRET measurements needed for reliable segregation of models is reported in the "#pairs" column. Initially predicted FRET pairs are used for guiding, while an extended set of FRET pairs is used for cross-validation.

(**)*The starting ensembles, FRET networks and optimization cycles are summarized for all proteins in* **Supplementary Fig. 7**.

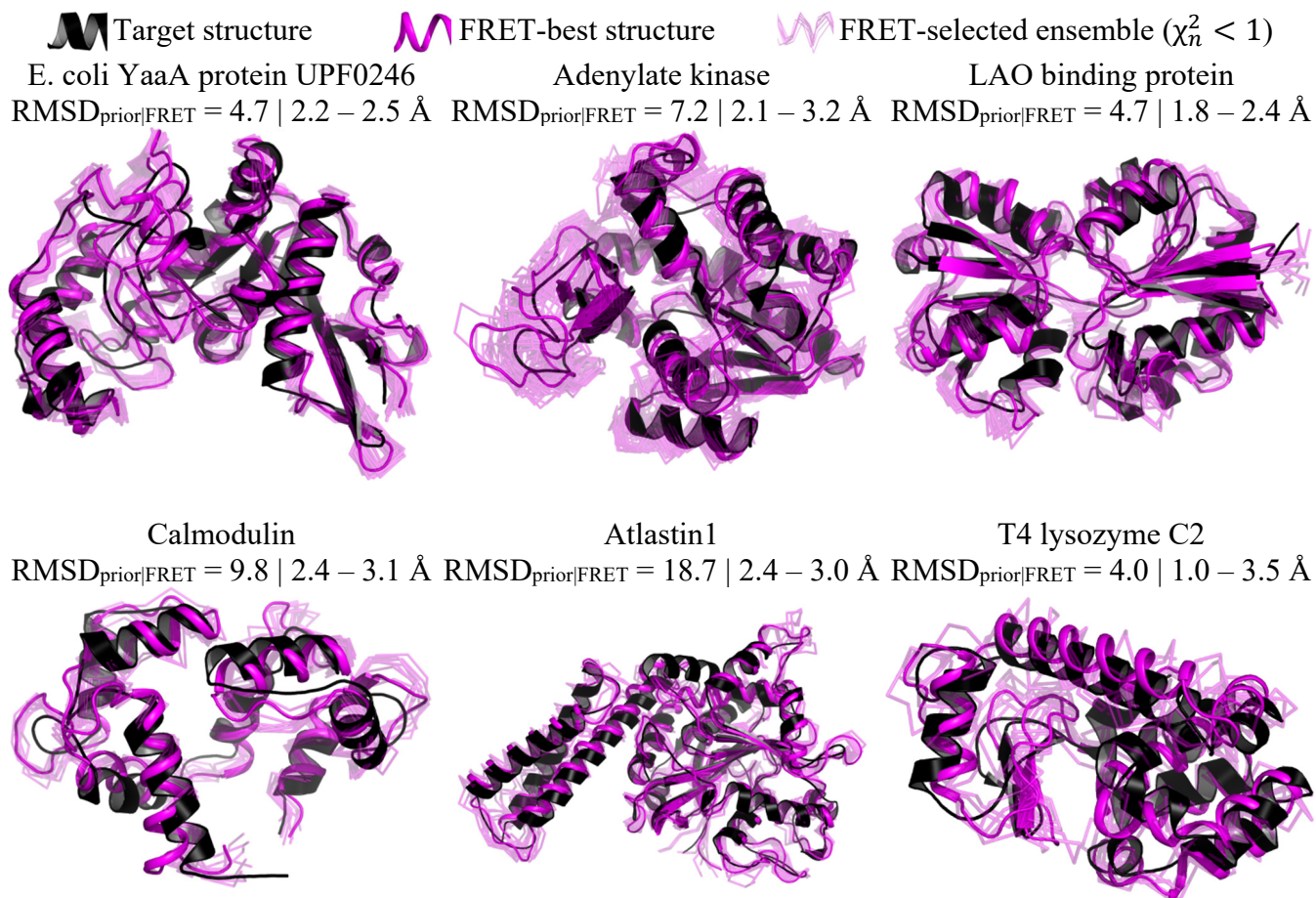

**Figure 2** | *Structures obtained by FRET-assisted modeling (magenta) and target X-ray structures (black) are shown for each of the benchmarked proteins. FRET-selected structures are depicted in transparent magenta as a measure for precision; a confidence level of 68% is assumed.*

In summary, we demonstrate against simulated and experimental data that accurate, efficient, and largely automated protein structure determination is possible based on optimally designed FRET experiments and structural modeling at multiple scales. In our view, the obtained results provide a major step ahead for quantitative FRET-assisted structural modeling. Furthermore, the approach described here should also be applicable to other label-based techniques such as EPR, paramagnetic relaxation enhancement NMR, vibrational spectroscopy, and their combinations, with minimal changes to the implementation.

### Acknowledgments

This work was funded in part by the German Research Foundation (DFG) within the Collaborative Research Center SFB 1208 "Identity and Dynamics of Membrane Systems – From Molecules to Cellular Functions" (TP A03 to HG and TP A08 to CS) and by the European Research Council through the Advanced Grant 2014 (number 671208) to CS. HS acknowledges support of the Alexander von Humboldt, NSF-BIO 1749778. We are grateful for computational support and infrastructure provided by the "Zentrum für

Informations- und Medientechnologie" (ZIM) at the Heinrich Heine University Düsseldorf and the computing time provided by the John von Neumann Institute for Computing (NIC) to HG on the supercomputer JURECA at Jülich Supercomputing Centre (JSC) (user ID: HKF7). Authors are grateful to Christian Hanke for the help with the preparation of data files for submission to PDB-Dev.

### Author contributions

M.D. developed methods and performed programming and computations. H.S., D.R., K.H. and T.P. designed T4L FRET network, prepared, labeled and analyzed experimental data for T4 Lysozyme. M.D. C.S. and H.G. analyzed data, discussed the results, and wrote the paper. C.S. and H.G. performed study design and supervised the project.

### Data availability

The original experimental data supporting the findings in this work are available from the corresponding authors upon reasonable request. Before publication, they will be uploaded to Zenodo. All structural models based on experimental FRET restraints will be deposited at PDB-dev using the **FLR-dictionary** extension (developed by PDB and the Seidel group) on the IHM working group GitHub site. All structural models based on simulated FRET restraints will be deposited at the Model archive.

### Code availability

Our automated FRET workflow relies upon three software tools:
1. Olga [1] software for FRET-screening and optimal FRET pair selection (experiment planning)
2. NMSim webserver [2] – a coarse-grained geometric simulations software for unrestrained conformational sampling and FRET-guided coarse-grained modelling.
3. FRETrest [3] and LabelLib [4] – a set of command-line tools for FRET-restrained molecular dynamics simulations

NMSim and Labelib are already accessible publicly. To provide early access to all of our software tools for the reviewers and editors, we created special github account, that is given access to the repositories of Olga and FRETrest.

Github username: mpc-guest
password: Rh0damine110

The software and documentation are already accessible. Currently we continue improving the documentation. Before publication all components will be made public on github

[1]: https://github.com/Fluorescence-Tools/Olga
[2]: http://nmsim.de
[3]: https://github.com/Fluorescence-Tools/FRETrest
[4]: https://github.com/Fluorescence-Tools/labellib

# Online Methods

1. **Proteins used in the benchmark.**

To demonstrate our structural modeling approach and assess its performance, we selected six protein systems: LAO binding protein, adenylate kinase, calmodulin, atlastin1, *E. coli* YaaA protein and T4 lysozyme. The proteins were selected such that the conformational transition between *prior* and target conformer covers different types of internal motions (hinge-bending, sheer, bend, and twist). The proteins span a wide range of sizes, from 148 amino acids (calmodulin) to 409 amino acids (atlastin1).

For each of the first four proteins[23-32], at least two crystal structures are known. One crystal structure is considered the "target" structure, another one of a different conformational state was used as a *prior* (**Table 4**). *E. coli* YaaA protein was one of the targets of the CASP11 experiment[14] (T806). For this protein, 10 homology models provided by the participants of the CASP11 experiment were used as the *prior*. These seed structures were selected from 639 complete protein models submitted to the CASP11 experiment by, first, removing those structures that are similar to the target ($C_\alpha$ atom RMSD < 4.6 Å). The remaining 589 models were clustered into 100 clusters by their secondary structure using Hierarchical agglomerative clustering[33]. From these 100 cluster representatives, 10 were selected by hand such that they represent different tertiary structures and different $C_\alpha$ atom RMSD with respect to the target (4.6 ≤ RMSD ≤ 14.6 Å, CASP model ID: Tc806TS041_1, Tc806TS065_1, Tc806TS276_1, Tc806TS345_1, Tc806TS357_1, Tc806TS420_1, Tc806TS428_1, Tp806TS065_1, Ts806TS065_1, Ts806TS276_1).

2. **Quality metric for evaluation of FRET pair sets: $\langle\langle RMSD\rangle\rangle$.**

To assess, how well certain sets of DA pairs can help resolving a protein structure, we introduce a quality parameter $\langle\langle RMSD_{\#conf}\rangle_{\#ref}\rangle$, or short $\langle\langle RMSD\rangle\rangle$, an estimate for expected precision (uncertainty). Assuming that one unknown structure from the *prior* ensemble is correct, $\langle\langle RMSD\rangle\rangle$ is defined to serve as an estimate for what would be the precision if we determined it from experiment (**Supplementary Fig. 8**). Conceptually, first, we take an arbitrary reference model from the prior and assume that it corresponds to the 'true' structure of the molecule in experiment. For this reference, a full reference set of FRET observables is simulated. Second, FRET observables are simulated for each conformer in the *prior* and tested against the reference set of observables, $\chi^2$ and *p*-values are calculated, and the precision $\langle RMSD_{\#conf}\rangle$ for this reference is determined. This procedure is repeated for each reference conformer from the prior, and the average over $\langle RMSD_{\#conf}\rangle$ is calculated, yielding $\langle\langle RMSD_{\#conf}\rangle_{\#ref}\rangle$ (**Supplementary Fig. 8**).

For any given set of *N prior* conformations, $\langle\langle RMSD\rangle\rangle$ is calculated in three stages (**Supplementary Fig. 1**): First, a N x N matrix is formed from RMSD values of all pairwise combinations of conformers in the prior:

$$RMSD_{conf,ref} = \sqrt{\frac{1}{N_{atoms}} \sum_{at=1}^{N_{atoms}} \left\| \overrightarrow{r_{ref,at}} - \overrightarrow{r_{conf,at}} \right\|^2} \qquad (1)$$

*ref* stands for the reference conformer, $conf$ for the tested conformer, $\overrightarrow{r_{at}}$ is the position of an atom in space, $N_{atoms}$ is the number of atoms in the protein. In this study only $C_\alpha$ atoms are considered for RMSD estimation. Second, the *N* x *N* matrix of FRET *p*-values are calculated for the same conformer pairs. To evaluate *p*-values, we start by calculating $\chi^2_{conf,ref}$, the $\chi^2$ of a tested conformation with respect to the reference conformation:

$$\chi^2_{conf,ref} = \sum_{i=1}^{N_{measurements}} \left(\frac{R^{(i)}_{conf} - R^{(i)}_{ref}}{\delta^{(i)}_{ref}}\right)^2 \qquad (2)$$

$R^{(i)}_{conf}$ is the FRET distance calculated for FRET pair $i$ on a conformational model $conf$, $R^{(i)}_{ref}$ is the corresponding distance in the reference conformer, $\delta^{(i)}_{ref}$ is the expected experimental error. $N_{dof}$ is the number of degrees of freedom in $\chi^2$ test:

$$N_{dof} = N_{measurements} - N_{SMP} \qquad (3)$$

$N_{measurements}$ is the number of FRET measurements (pairs) taken, $N_{SMP}$ is the number of independent relevant coordinates (parameters) for the conformational model (see below). For every conformer pair, we can calculate a *p*-value or a probability that a sample $\chi^2$ will be larger than $\chi^2_{conf,ref}$:

$$p_{conf,ref} = p(\chi^2_{conf,ref}, N_{dof}) = \int_{\chi^2_{conf,ref}}^{+\infty} f_{N_{dof}}(\chi^2)\, d\chi^2 \qquad (4)$$

$f_{N_{dof}}(\chi^2)$ denotes the chi-squared distribution:

$$f_{N_{dof}}(\chi^2) = \frac{1}{2^{N_{dof}/2}\Gamma(N_{dof}/2)}(\chi^2)^{N_{dof}/2-1}e^{-\chi^2/2} \qquad (5)$$

$\Gamma$ is the Gamma function. Third, $\langle\langle RMSD \rangle\rangle$ is evaluated as a weighted average over the RMSD matrix using the respective *p*-values as weights. $\langle\langle RMSD \rangle\rangle$ is a double average over all reference conformers as well as all conformers being tested:

$$\langle\langle RMSD \rangle\rangle = \frac{1}{N_{conf}} \sum_{ref=1}^{N_{conf}} \frac{\sum_{conf=1}^{N_{conf}} p_{conf,ref} RMSD_{conf,ref}}{\sum_{conf=1}^{N_{conf}} p_{conf,ref}} \qquad (6)$$

3. **FRET screening.**

To assess, how well a given structural model or structural ensemble agrees with experimental FRET data, we calculate the $\chi^2$ value for each structure in the ensemble. To do that, we need to estimate FRET observables corresponding to the specified conformer. We achieve this by simulating the Accessible Volume (AV) of the fluorophore attached to a protein by a flexible linker[12] (see **Supplementary Note 5**).

In general, reduced chi-squared $\chi^2_r$, also known as chi-squared per degree of freedom, is used as an absolute quality parameter of a model:

$$\chi^2_r = \chi^2/N_{dof} \qquad (7)$$

However, for values of $N_{dof} < 30$, a constant confidence level corresponds to different values of $\chi^2_r$. Therefore, using $\chi^2_r$ to compare models with different $N_{dof}$ is inconvenient. To overcome this, we introduce an alternative metric, normalized chi-squared $\chi^2_{n,68\%}$, which equals to 1 for $p = 68\%$ (one sigma) by definition, independent of the $N_{dof}$ value (**Supplementary Fig. 9**):

$$\chi^2_n = \chi^2_{n,68\%} \equiv \chi^2/Inv.\chi^2(p=0.68, N_{dof}) \qquad (8)$$

Where

$$Inv.\chi^2(p, N_{dof}) = \frac{2^{-N_{dof}/2}}{\Gamma(N_{dof}/2)} p^{-N_{dof}/2-1} e^{-1/(2p)} \qquad (9)$$

is the inverse chi-squared distribution. To visualize the precision of the generated structural ensembles, we display conformations on two-dimensional plots (**Supplementary Fig. 2**).

Given an ensemble of structural models, $\chi_n^2$ can be calculated for each conformer. Structures that show better agreement with FRET data have lower $\chi_n^2$. If the FRET-selected ensemble ($\chi_n^2 < 1$) is too diverse (e.g., $RMSD_{ij} > 3$ Å), extra FRET pairs can be selected and measured to improve resolution (see below). In this benchmark reference FRET data ($R_{ref}^{(i)}, \delta_{ref}^{(i)}$) were determined from experiment for T4 lysozyme and simulated for other benchmarked proteins using the 'true' crystal structure conformations, as described previously[8]. Structures of T4 lysozyme and its homologs from the PDB were screened against the experimental datasets C1 and C2 in order to select reference conformations for each state (**Supplementary Fig. 5**). As a result PDB ID 172L appears to correspond to C1, and PDB ID 3GUN was selected for C2.

4. **Selection of a set of optimal FRET pairs**.

To maximize the precision of FRET-assisted protein structure determination under the condition of a limited number of experimental measurements, we developed a method for automated determination of the most informative labelling sites and donor-acceptor (DA) pairs. We define sets of pairs to be most informative if they lead to the highest expected precision, i.e., lowest ⟨⟨RMSD⟩⟩, of a structural model. To find such an optimal DA pair set, we test three different feature selection algorithms (**Supplementary Fig. 10**): greedy forward selection (**Supplementary Note 2**), greedy backward elimination (**Supplementary Note 3**), and an algorithm based on mutual information and inspired by a Minimum Redundancy Maximum Relevance (mRMR) algorithm[34] (**Supplementary Note 4**). FRET pairs are selected among the full set of all possible pairwise combinations of available labeling sites. Labeling sites can be excluded from calculations based on additional prior information provided by the user, e.g. accessibility, chemical nature and influence on function and stability as determined from mutation analysis or sequence coevolution data. For the proof of principle study with simulated data, we assume that these effects are negligible. However, considering the experimental data sets of T4L, care was taken to avoid these problems. For T4L automated FRET pair selection was performed from only 33 FRET pairs as opposed to theoretically possible $162^2/2$ residue-residue combinations. These 33 pairs were earlier chosen by authors for a functional study of T4L[21] (see **Online Methods section 9**). Despite of this low number of available FRET-pairs, only minor decrease in expected precision was observed as compared to other proteins (**Supplementary Fig. 7**).

In greedy forward feature selection, in the first iteration, ⟨⟨RMSD⟩⟩ is calculated for each possible DA pair, and that pair is selected for the DA set that yields the minimal ⟨⟨RMSD⟩⟩. In the next iterations, DA pairs remaining from the previous iteration are probed against the DA set to determine which one leads to the largest decrease in ⟨⟨RMSD⟩⟩; that DA pair is then added to the DA set. The algorithm stops when a desired ⟨⟨RMSD⟩⟩ is reached. Therefore, for conformational ensembles < 100,000 structures, the current implementation converges in less than a day on a 4-core desktop computer.

In greedy backward elimination, in the first iteration, ⟨⟨RMSD⟩⟩ is calculated for DA sets containing all possible DA pairs but one. That pair is eliminated for which the remaining DA set yielded the smallest ⟨⟨RMSD⟩⟩; the remaining DA set is narrowed further in an iterative manner. The algorithm needs to run as many iterations as there are DA pairs available and is therefore slower than the greedy selection algorithm. One run of this algorithm for an ensemble of less than 10,000 conformers completes in about one day on a 4-core desktop computer in the current implementation.

In the mutual information-based DA pair selection algorithm Shannon conditional entropies are calculated for all pairwise combinations of DA pairs. In the first iteration, the DA pair with the highest Shannon entropy is selected. In the next iterations, the DA pair with the highest minimum Shannon conditional entropy with respect to the previous DA pairs is selected (**Supplementary Note 4**). That way, the DA pair providing the highest amount of new information not provided by the previously selected DA pairs is selected. One run of this algorithm for an ensemble of less than 100,000 conformers completes in about one day on a 4-core desktop computer in the current implementation.

5. **Estimation of the complexity of the structural model.**

Estimation of complexity for a structural model that is used in integrative protein structure determination is essential for quantitative accuracy assessment and automated experiment design. We quantify the complexity of a structural model by the number of relevant independent parameters (coordinates, $N_{SMP}$) needed to describe the corresponding conformational ensemble to a certain precision $\langle\langle RMSD\rangle\rangle$. If the structural model is simple, $N_{SMP}$ can be calculated analytically, for example, for a rigid body model, $N_{SMP} = (N_{bodies}-1) * 6 - N_{bonds}$. For non-rigid body models, coming from other computational tools, an analytical expression for $N_{SMP}$ is usually unavailable. Examples of such tools are numerous: molecular dynamics simulations (all-atom or coarse-grained), normal mode-based models, homology models, elastic network models, and others.

We thus introduce a heuristic approach for automated $N_{SMP}$ determination, which requires as an input only the user-provided conformational ensemble. Initially, to obtain an $N_{SMP}$ estimate, we start by assuming $N_{SMP,0} = 0$, and determine a set of DA pairs needed to describe the conformations within an ensemble with a desired precision $\langle\langle RMSD\rangle\rangle$ employing our DA pair selection algorithm. Each DA pair can be seen as a coordinate, and the number of DA pairs corresponds to our definition of $N_{SMP}$. Second, we use the number of FRET pairs as predicted by the algorithm at the first stage as the true $N_{SMP}$ and re-run the pair selection to obtain an estimate for the number of measurements needed for FRET-based structure determination. Thereby, the number of required measurements is always larger than the model's complexity ($N_{SMP}$), reflecting that statistical significance can only be properly assigned to an overdetermined model ($N_{dof} > 0$, see eq. 3).

For a FRET-restrained structural model (e.g., generated by FRET-guided NMSim or FRET-restrained MD simulations, see below) the same procedure can be used. Presuming that the explored degrees of freedom in the FRET-restrained model cover all FRET restraints, one can conservatively assume $N_{SMP} \geq N_{FRET\ restraints}$. In this study, we use $N_{SMP} = N_{FRET\ restraints}$ as a complexity estimate for all FRET-restrained models. Hence, FRET-guided structural sampling must be followed by an additional round of pair selection, so that more FRET pairs are measured for cross-validation.

Overall, these approximations apparently lead to good $N_{SMP}$ estimates, and further independent measurements do not change $\chi_n^2$ significantly (**Fig. 1i**). Reliability of $N_{SMP}$ estimates is also evident from the correlation between $\chi_n^2$ and accuracy against the target structure (**Supplementary Fig. 6**).

6. **Unbiased conformation sampling by NMSim.**

Structural ensembles unbiased by experimental FRET data were generated by the NMSim software[13]. Ten independent and unbiased NMSim simulations generating 10,000 conformations each were performed, starting from the *prior* structure and using default parameters for sampling of large scale motions. These trajectories are clustered and serve as *prior* candidates. NMSim is a normal mode-based geometric simulation approach for multiscale modeling of protein conformational changes that incorporates information about preferred directions of protein motions into a geometric simulation algorithm. NMSim follows a three-step protocol: In the first step, the protein structure is coarse-grained by the software FIRST[35] into rigid parts connected by flexible links[36]. In the second step, low-frequency normal modes are computed by rigid cluster normal mode analysis (RCNMA)[37]. In the third step, a linear combination of the first 10 normal modes was used to bias backbone motions along the low-frequency normal modes, while the side chain motions were biased towards favored rotamer states. Detailed list of used simulation parameters is given in the **Supplementary Note 6**.

### 7. FRET-guided NMSim.

To improve the sampling of the conformational space in regions most relevant according to experiment, we extended the NMSim approach by a Markov Chain Monte Carlo step to prioritize conformations lying in such regions (**Supplementary Fig. 3**). In every NMSim iteration, the generated conformation is scored with respect to its agreement with experimental data using the $\chi_n^2$ metric. Then, according to the Metropolis-Hastings approach,

$$p_{accept} = \exp\left(\frac{\chi_{n,previous}^2 - \chi_{n,current}^2}{kT}\right) \quad (10)$$

is computed, and the conformation is accepted and used in the next NMSim iteration if $p$ is larger than a uniformly distributed random number sampled from the range [0, 1]; else, the conformation is discarded, and the previous one is used in the next NMSim iteration. As a result, conformations are generated that are both stereochemically plausible and agree with experimental data. To improve the sampling and enable the exploration of multiple local minima, an annealing procedure is applied in which $kT$ is varied from almost 0 to 1 units of $\chi_n^2$ and back to 0 (see **Supplementary Note 6**). A single FRET-guided NMSim simulation contains two such annealing cycles. If, models with good FRET agreement ($\chi_n^2 \to 1$) cannot be obtained from FRET-guided simulations, alternative seed structures should be considered.

### 8. FRET-restrained MD.

To reconstruct structures to maximum detail, we developed a procedure to incorporate FRET-restraints in atomistic molecular dynamics (MD) simulations (**Supplementary Fig. 4**). To generate the restraints, first Accessible Volume (AV) calculations are performed for each labeling position. Second, pseudo atoms are positioned at the mean position of every accessible volume. These pseudo atoms do not interact with protein or solvent atoms. To keep the pseudo atoms in their initial positions relative to the corresponding part of the back-bone, harmonic restraints are used: Pseudo bonds are created between the pseudo atom and $C_\alpha$ and $C_\beta$ atoms of amino acids up to two residues towards the C- or N-termini of the protein from the amino acid, where the fluorophore linker is attached. Thus, each pseudo atom is anchored to ten nearby backbone atoms. The positions of pseudo atoms, the anchoring bonds, and FRET restraints are recalculated every 2 ns during the simulation to account for changes in local structure.

To mimic the measured FRET distances, pseudo atoms are restrained with respect to each other using harmonic-linear restraints. If the distance between pseudo atoms corresponds exactly to the measured donor-acceptor distance, no additional force is applied to pseudo atoms. To prevent unphysical unfolding of the protein, the FRET-restraint force is capped at an empirically determined value $F_{max} = 50$ pN, which is reached when the bond length ($R_{DA}$) is more than one standard error ($\delta_{exp}$) away from the optimum ($R_{exp}$, **Supplementary Fig. 4D**). The error for each FRET distance is determined from experimental data. Force constants for each FRET-restraint are tuned such that for every pseudo atom the magnitude of the total FRET-restraints vector is $\leq F_{max}$, resulting in force constants for FRET restraints in the range from 0.7 to 14 pN / Å, depending on their collinearity. Force constants of the pseudo bonds that attach pseudo atoms to their local backbone atoms are set 10 times higher than those for FRET restraints. FRET restraints are implemented using the AMBER interface for NMR restraints ("DISANG" file).

It is worthwhile to note that, unlike the immediate position of a fluorophore, its *mean* position with respect to the local backbone does not change as quickly. This way, we avoid complications of explicit dye simulations, such as potential inaccuracies of dye force field parametrizations and large convergence times (> 100 ns[38]) of fluorophore diffusion. Furthermore, if FRET restraints were applied to explicitly modelled fluorophores directly, the flexible dye linker would become an entropic spring[39] and absorb virtually all of the strain. Finally, FRET observables determined in experiment have a statistical nature: they represent state-specific ensemble averages and underlying distributions, rather than immediate quantities. Therefore, application of 'statistical' FRET restraints to pseudo atoms that are constructed to mimic statistically averaged fluorophore positions is more straightforward and effective.

The AMER16 suite of molecular simulation codes[40] was used to perform MD simulations. All co-crystallized waters and ligands were removed from the crystal structures. Hydrogen atoms were removed and re-added by tleap[41] from the AMBER Tools suite. The TIP3P explicit water model[42] was used to solvate proteins in a periodic truncated octahedral box with at least 12 Å of solvent in every direction from the protein surface. Sodium and chloride counter ions were added to neutralize the systems. MD simulations were performed with the ff14SB force field[43] using the GPU version of pmemd[44]. The SHAKE algorithm[45] was used to constrain bond lengths of hydrogen atoms. Long-range electrostatic interactions were evaluated using the particle mesh Ewald method[46]. Hydrogen mass repartitioning[47] and a time step of 4 fs were used. A five-stage equilibration procedure was pursued: First, 100 steps of steepest descent and 400 steps of conjugate gradient minimization were performed, while solute atoms were restrained to their initial positions by harmonic restraints with force constants of 5 kcal mol$^{-1}$ Å$^{-2}$. Second, the temperature of the system was raised from 100 K to 300 K in 50 ps of NVT-MD simulations. Third, 150 ps of NPT-MD simulations were performed to adjust the system density. Finally, the force constants of harmonic restraints were gradually reduced to zero during 2 ns of NVT-MD simulations. Production NVT-MD simulations were carried out at 300 K, using the Berendsen thermostat[48] and a coupling constant of 0.5 ps. Three independent replicas of MD simulations (1µs per simulation) were performed for each system using different random number generator seeds to assign initial velocities.

### 9. T4 Lysozyme site specific mutation, purification and labeling.

T4L site directed mutagenesis was performed on the cysteine-less pseudo-wild-type encoded backbone using the pET11a (Life Technologies, Corp) vector as previously described[49-51]. For protein expression and purification, the plasmid containing T4L desired mutations (a unnatural amino acid –p-acetyl-L-phenylalanine or pAcPhe, in the N-terminal subdomain (NTsD) and the replacement to a Cys in the C-terminal subdomain (CTsD)) was co-transformed with pEVOL[50] for the introduction of (pAcPhe) into BL21(DE3) *E. coli* strains (Life Technologies Corp.). Transformed *E. coli* were plated onto LB- agar plates supplemented with ampicillin and chloramphenicol for single colony selection. For each variant, a single colony was inoculated into 100 mL of LB with antibiotics and grown overnight at 37 °C in a shaking incubator, followed by inoculation of a 1 L LB medium supplemented with the respective antibiotics and 0.4 g/L of pAcPhe (SynChem) with 50 mL of the overnight culture. The culture was grown at 37 °C until an OD600 of 0.5 was achieved, for further induction. The protein production was induced for 6 hours by addition of 1 mM IPTG and 4 g/L of arabinose. Harvested cells were lysed in 50 mM HEPES, 1 mM EDTA, and 5 mM DTT pH 7.5 and purified using a monoS 5/5 column (GE Healthcare) with an eluting gradient from 0 to 1 M NaCl according to standard procedures. High-molecular weight impurities were removed by passing the eluted protein through a 30 kDa Amicon concentrator (Millipore), followed by subsequent concentration and buffer exchange to 50 mM PB, 150 mM NaCl pH 7.5 of the protein flow through with a 10 kDa Amicon concentrator.

Site specific labeling of T4L was accomplished using orthogonal chemistry following manufacturer suggestion. For labeling the Keto functional group of pAcPhe at the NTsD, the Alexa 488 with hydroxylamine linker chemistry was used (Life Technologies Corp.). Cysteine sites were labeled via a thiol reaction with maleimide linkers of Alexa-647. FRET or DA variants were labeled sequentially - first thiol and second the keto handle[51]. A proper Donor Only reference sample was only kept before proceeding with the acceptor labeling. The selected FRET pair has a Förster distance $R_0$ of 52 Å.

### 10. FRET Experiments and Analysis.

To resolve the conformational heterogeneity of T4L, Donor only and FRET labeled T4L variants were studied by time-resolved fluorescence spectroscopy using Time Correlated Single Photon Counting (TCSPC) and single-molecules studies with confocal multiparmter fluorescence detection.

Donor only and FRET labeled T4L variants were measured by TCSPC using either an IBH-5000U (IBH, Scotland) or a Fluotime 200 (Picoquant, Germany) system. The excitation source of the IBH machine were a 470 nm diode laser (LDH-P-C470, Picoquant, Germany) operating at 10 MHz for donor excitation and a 635 nm (LDH-P-C635, Picoquant, Germany) for acceptor excitation. The excitation and emission slits were set to 2 nm and 16 nm, respectively. The excitation source of the Fluotime200 system was a white light laser (SuperK extreme, NKT Photonics, Denmark) operating at 20 MHz for both donor (485 nm) and acceptor (635 nm) excitation with excitation and emission slits set to 2 nm and 5 nm, respectively. Additionally, in both systems, cut-off filters were used to reduce the amount of scattered light (>500 nm for donor and >640 nm for acceptor emission).

For green detection, the monochromator was set to 520 nm and for red detection to 665 nm. All measurements were conducted under magic angle conditions (excitation polarizer 0°, emission polarizer 54.7°, VM), except for anisotropy where the position of the emission polarizer was alternately set to 0° (VV) or 90° (VH).

In the IBH system, the TAC-histograms were recorded with a bin width of 14.1 ps within a time window of 57.8 ns, while the Fluotime200 was set to a bin width of 8 ps within a time window of 51.3 ns. The average number of collected photons per sample were $30 \times 10^6$ photons.

A global joint analysis of the donor only and FRET labeled samples was implemented in order to assure proper donor reference samples, determination of the mean inter-dye distances, $\langle R_{DA} \rangle$, and assignment of states by sharing the population parameters on the FRET labeled samples. The analysis and justification of the methods are reported in Sanabria et al[21]. In short, the donor only labeled samples were fit with a multiexponential decay model (eq. 25, Peulen et al)[52]. All FRET induced donor VM decays were fit using the corresponding donor only decay parameters with a sum of Gaussian distributed states to derive $\langle R_{DA} \rangle$. By using a global analysis, we assure conformational states are assigned via the linked population fractions. A 2σ statistical uncertainty and an error propagation rule considering $\kappa^2$ error was used to consider the overall uncertainty (+/- err). The derived distances for two states are presented in in **Supplementary Table 1**. The error estimation considers: (*i*) upper estimates for the uncertainty of the orientation factor[53], $\kappa^2$, (*ii*) statistical uncertainties of the analysis[52], (*iii*) estimates for systematic errors due to imprecise reference samples[52], and (*iv*) uncertainty estimates for modelling the spatial distribution of the dyes based on the dye's residual anisotropies[9] (see **Supplementary Table 2**).